\documentclass{jfm}
\usepackage{graphicx}
\usepackage{epstopdf, epsfig}

\shorttitle{Boundary layer of elastic turbulence}
\shortauthor{S. Belan, A. Chernykh and V. Lebedev}

\title{Boundary layer of elastic turbulence}

\author{S. Belan\aff{1}
  \corresp{\email{sergb27@yandex.ru}},
  A. Chernykh\aff{2,3}
 \and V. Lebedev\aff{4,5}}

\affiliation{\aff{1}Massachusetts Institute of Technology, Cambridge, Massachusetts 02139, USA
\aff{2}Institute of Automation and Electrometry SB RAS, 630090, Academician Koptyug ave. 1, Novosibirsk, Russia
\aff{3}Novosibirsk State University,  630073,  Prospekt K. Marksa 20, Novosibirsk, Russia
\aff{4}Landau Institute for Theoretical Physics RAS, 142432, Ak. Semenova 1-A,
Chernogolovka, Moscow region, Russia
\aff{5}Higher School of Economics, 101000, Myasnitskaya 20, Moscow, Russia}

\begin{document}

\maketitle

\begin{abstract}
We investigate theoretically the near-wall region in elastic turbulence of a dilute polymer solution in the limit of large Weissenberg number. As it was established experimentally, elastic turbulence possesses a boundary layer where the fluid velocity field can be approximated by a steady shear flow with relatively small fluctuations on the top of it.
Assuming that at the bottom of the boundary layer the dissolved polymers can be considered as passive objects, we examine analytically and numerically statistics of the polymer conformation, which is highly nonuniform in the wall-normal direction. Next, imposing the condition that the passive regime terminates at the border of the boundary layer, we obtain an estimate for the ratio of the mean flow to the magnitude of flow fluctuations. This ratio is determined by the polymer concentration, the radius of gyration of polymers and their length in the fully extended state.
The results of our asymptotic analysis  reproduce the  qualitative features of elastic turbulence at finite Weissenberg numbers. 
\end{abstract}


\section{Introduction}

Polymer solutions attract much experimental and theoretical attention which is naturally explained by their fascinating non-Newtonian behavior \citep{Bird_I}. 
Dissolving even small amount of polymers in an ordinary fluid may dramatically change its hydrodynamic and rheological properties due to appearance of elastic degrees of freedom. 
One of the most striking manifestation of the non-Newtonian dynamics is a chaotic fluid motion which is observed at low Reynolds number $\Rey$ exhibiting three main features \citep{GS00,GS_2001,GS_2004,GCS05,BSS06_1,BSS06_2}: pronounced growth in flow resistance, algebraic decay of velocity power spectra over a wide range of scales, and orders of magnitude more efficient mixing than in an ordered flow. Since these properties are analogous to those of hydrodynamic turbulence, the chaotic state of the polymer solution has been named elastic turbulence.

Despite the close similarity between hydrodynamic turbulence and elastic turbulence, physical mechanisms that underlies the two kinds of random motion are different. The former is known to occur at large enough $\Rey$ due to instabilities arising from the non-linear inertia term in the Navier-Stokes equation. In contrast, elastic turbulence takes place at vanishingly small $\Rey$ where the effects of fluid inertia play no role.
The main source of instabilities  leading to elastic turbulence is the  elastic stresses created by the polymer stretching in the flow. The transition from laminar flow to elastic turbulence is controlled by the so-called Weissenberg number $\textit{Wi}$, that is a ratio of the polymer linear relaxation time to the characteristic time of the flow dynamics. At large value of $\textit{Wi}$, the majority of the dissolved polymers are above the coil-stretch transition that guarantees their strong back-reaction on the fluid motion provided the polymer concentration is high enough \citep{GCS05,Linear1,Chertkov_2000,Linear2}.

The first experiments on elastic turbulence have used three flow geometries with curved streamlines \citep{GS00,GS_2001,GS_2004}: von Karman swirling flow between two disks, Couette-Taylor flow between cylinders, and Dean flow in a curvilinear channel. 
Recently,  purely elastic instabilities have been observed experimentally in a straight channel \citep{Pan_2013,Bodiguel_2015} and elastic turbulence has been demonstrated numerically for the viscoelastic Kolmogorov flow \citep{Berti_2008,Berti_2010}. 
Although the streamline curvature is not a crucial ingredient, it allows to reduce the critical Weissenberg number for the instability onset.

In which aspects does elastic turbulence considerably differ from its inertial counterpart due to the aforementioned difference in the sources of flow instability?
Although the phenomenon of elastic turbulence is known for two decades, this issue still remains poorly understood.
The present paper demonstrates that at least one such aspect is the relation between the mean and the fluctuating components of the flow in the boundary layer region. The general rule for hydrodynamics turbulence states that the typical magnitude of random motion is of the order of the mean velocity variation at the considered spatial scale \citep{LL6}. 
In the context of wall-bounded turbulence, this means that fluctuations in the viscous boundary sublayer are of the order of the mean flow velocity.
As it is shown below, this rule becomes completely wrong in the case of wall-bounded elastic turbulence.

Polymer solutions can be characterized at two distinct levels: macroscopic and microscopic. The macroscopic approach treats the system as a continuous medium and, thus, focuses on the bulk parameters, which are averages of microscopic variables over the scales much larger than the inter-polymer distance. In particular, the macroscopic description of the fluid velocity field is based on the generalized Navier-Stokes equation incorporating the elastic stresses \citep{Linear1,Balkovsky_2001,Linear2}. At the microscopic level one deals with individual polymer molecules advected by the fluid flow.
The microscopic approach has been extensively used to study the dynamics and statistics of polymer conformation  \citep{CKLT05,Celani_2005a,Celani_2005b,Afonso_2005,Tur07,Vincenzi_2015,Ahmad_2016}. In this paper we consistently implement both microscopic and macroscopic approaches to understand the properties of the boundary layer of  elastic turbulence.

The emergence of a boundary layer where the fluid velocity field is approximated by a steady shear flow with relatively small fluctuations has been revealed in the experimental studies of elastic turbulence in different flow geometries \citep{BSS06_1,BSS06_2,Jun_2011}.
From the theoretical point of view, the boundary layer is the region immediately adjacent to the wall within which the viscous stresses in a fluid are much larger than the elastic stresses associated with hydrodynamic stretching of polymers. 
Qualitatively, suppression of elastic stresses near the wall occurs because the gradient of the wall-normal component of the random velocity vanishes at the wall due to the non-slipping boundary condition. 
The viscosity smooths the fluid velocity field which still remains a random function of time due to the fluctuations induced by the bulk flow where elastic stresses are relevant.
A high-$\Rey$ analogue of the boundary layer in the Newtonian fluids is the well-known viscous boundary sublayer defined by the condition that the viscosity dominates over the fluid inertia \citep{LL6}.

Let us expose the logic of our analysis.
Since the backward reaction of the polymers on the flow diminishes at the bottom of the boundary layer, the polymers passing near the wall can be considered as passive. That dictates peculiarities of the velocity field structure and allows us to describe the spatially non-uniform statistics of the polymer conformation which is determined by interplay of the average shear flow and velocity fluctuations.
Next we use this information to find the elastic stresses responsible for the polymer back reaction on the flow. This back reaction is negligible in comparison with the viscous forces at small distances from the wall, but it eventually come into game at larger scales. From the condition that the elastic stresses become comparable with the viscous stresses at the border of the boundary layer, we extract the ratio of the mean flow to the magnitude of flow fluctuations.
At the end of the paper the applicability range of our theory is discussed.

\section{Velocity field in the boundary layer}

At considering the boundary-layer properties, we treat the wall as flat.
For the channel flow, it is correct if the thickness of the boundary layer $L$ is of the order or smaller than the channel radius and its curvature.
This condition also guaranties that correlation functions of the statistically stationary in time flow are homogeneous  along the wall.
Below, we consider the stationary case where the mean flow and the statistical properties of the flow fluctuations are time-independent.
Note that our analysis is also applicable to the near-disk boundary layer in the von Karman swirling flow provided we are interested in the region far enough from the rotation axis \citep{BSS06_1,BSS06_2}.

%
The Reynolds decomposition for the fluid velocity $\boldsymbol{v}$ reads
 \begin{equation}
 \label{decomposition}
 \boldsymbol{v}=\boldsymbol{U} +\boldsymbol{u},
 \end{equation}
where $\boldsymbol{U}(\boldsymbol{r})$ and $\boldsymbol{u}(\boldsymbol{r},t)$ are the mean and fluctuating parts of the flow velocity, respectively. Due to the dominant role of viscosity inside the boundary layer, the velocity field can be regarded as smooth and regular function of coordinates. In particular, the mean velocity is  approximated in the leading order by a shear flow 
\begin{equation}
\label{shear}
U_x = sz,
\end{equation}
where $z\ll L$ and $s$ is the shear rate. Here and below we choose a Cartesian reference system in such way that the $z$-axis is perpendicular to the wall, and the $x$-axis is directed along the velocity of the mean flow, see Fig. \ref{pic:orient}.

For the fluctuating velocity $\boldsymbol{u}$ the following proportionality laws are valid
 \begin{equation}
 u_{x,y}\propto z,\qquad
 u_z\propto z^2,
 \label{zdep}
 \end{equation}
provided $z \ll L$. The laws are consequences of the fluid velocity smoothness, the non-slipping boundary condition at the wall, and the incompressibility condition $\bnabla\bcdot\boldsymbol{u} = 0$. Note also that $\boldsymbol{u}$ varies along the wall at distances of the order of  $L$.
%
%

\section{Single polymer statistics}
In the dilute limit, we neglect the mutual interactions of the polymer molecules.
At $z\ll L$ a single polymer can be treated as a passive object  whose dynamics is governed by the equations
\begin{eqnarray}
\label{dynamics_r}
\partial_t{\boldsymbol{r}}=\boldsymbol{v}(\boldsymbol{r},t)+\boldsymbol{\xi}(t),\\
\label{dynamics_R}
\partial_t{\boldsymbol{R}}=-\gamma(R)\boldsymbol{R}+(\boldsymbol{R}\bcdot\boldsymbol{\nabla})\boldsymbol{v}(\boldsymbol{r},t)+\boldsymbol{\zeta}(t),
\end{eqnarray}
where $\boldsymbol{r}$ is the coordinate of the mass center of the polymer; $\boldsymbol{R}$ is the polymer end-to-end elongation vector; $\gamma(R)$  is the extension-dependent spring constant of the polymer; $\boldsymbol{\xi}$ and $\boldsymbol{\zeta}$ are the Langevin forces having zero means and time-decorrelated second moments $\langle \xi_i(t_1)\xi_j(t_2)\rangle =2\kappa\delta_{ij} \delta(t_1-t_2)$ and $\langle \zeta_i(t_1)\zeta_j(t_2)\rangle =2\eta\delta_{ij}\delta(t_1-t_2)$ with  $\kappa$ and  $\eta$ representing the diffusivity for translational and elongation degrees of freedom of polymer motion, respectively.

The physical picture behind Eqs. (\ref{dynamics_r}) and (\ref{dynamics_R}) is very simple: the polymer is advected along the Lagrangian trajectory being stretched by the velocity gradient and relaxing to its equilibrium shape due to elasticity.
Note that specific form of the function $\gamma(R)$ is not essential in the scope of this work.
The only information on the elastic properties of the polymer required for our analysis is the spring constant in the Hookean (linear) limit $\gamma_0=\gamma(0)$  and the maximum polymer extensibility $R_m$.
We imply that $R_m$ is small compared with the boundary layer thickness $L$.

In this section we use Eqs. (\ref{dynamics_r}) and (\ref{dynamics_R}) with the velocity field $\boldsymbol{v}$  given by Eqs. (\ref{decomposition}), (\ref{shear}) and (\ref{zdep}) to describe the statistics of the polymer conformation inside the boundary layer in its dependence on the distance from the wall $z$. Our analysis of is based on the several assumptions, to be justified: (i)  the Lagrangian correlation time $\tau_c$ of the flow in the near-wall region is much smaller than the characteristic time scale of  the polymer stretching  dynamics, (ii) the characteristic time scale of  the stretching  dynamics is much smaller than the characteristic time of polymer translational motion in the wall-normal direction, (iii) the thermal force $\boldsymbol{\zeta}$ can be neglected in comparison with the effect of velocity gradient stretching, and (iv) the polymer is strongly elongated along the mean velocity most of the time. A justification of these assumptions for the boundary layer region of high-$\it Wi$ elastic turbulence will be given below.

 \begin{figure}
  \centerline{\includegraphics[scale=1.08]{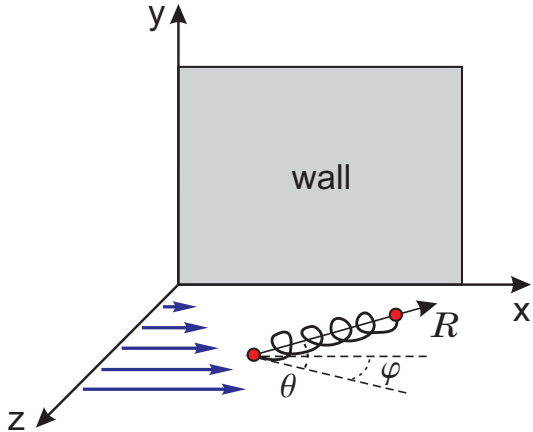}}
 \caption{Sketch of the polymer orientation
geometry.}
  \label{pic:orient}
 \end{figure}

Due to time scale separation (i) and (ii), the statistics of the polymer conformation at given spatial position is determined by the local intensity of the flow fluctuations which can be treated as shortly-correlated in time.
The conditions (iii) and (iv) in their turn mean that the main stochastic contribution to the polymer dynamics comes from the gradient of the $z$-component of the fluctuating velocity.
Then, one can apply the results of the theoretical works \citep{CKLT05,Tur07}, where exactly the same model assumptions have been exploited to examine the polymer dynamics in large mean shear with relatively weak flow  fluctuations on top of it.
For the sake of completeness we briefly reproduce the line of argumentations brought in these works.


The orientation of the polymer end-to-end vector $\boldsymbol{R}$ can be parametrized
by the spherical angles $\varphi$ and $\theta$, as shown in Fig. \ref{pic:orient}.
As was discussed in \citep{CKLT05}, in terms of these variables, Eq. (\ref{dynamics_R}) (with the Langevin
force omitted due to assumption (iii)) transforms into the following set of equations:
\begin{eqnarray}
\label{phi}
\partial_t \varphi=-s\sin^2 \varphi + \varsigma_\varphi,\\
\label{theta}
\partial_t \theta = -s\sin\varphi \cos\varphi \sin\theta\cos\theta +\varsigma_\theta,\\
\label{R}
\partial_t \ln R = -\gamma(R)+s\cos^2\theta\cos\varphi\sin\varphi +\varsigma_\parallel,
\end{eqnarray}
where $\varsigma_\varphi$, $\varsigma_\theta$ and $\varsigma_\parallel$  are random terms 
related to the fluctuating component of the fluid velocity.
Although these terms are spatially inhomogeneous, assumptions (i) and (ii) guarantee that statistics of the polymer elongation vector adjusts adiabatically to the local flow properties and, therefore, the internal degrees of freedom can be analysed separately from the translation motion.

The angular dynamics  described by Eqs. (\ref{phi}) and (\ref{theta}) is decoupled from the dynamics of the polymer length. 
Moreover, since the polymer is strongly elongated along the shear direction (assumption (iv)),  the orientation
angles are small most of the time, and  intensity of the random term $\varsigma_\varphi$ does not depend on $\theta$ in the leading order approximation.
This means that the angle $\varphi$ becomes decoupled from both $R$ and $\theta$.   
Treating $\varsigma_\varphi$ in Eq. (\ref{phi}) as a delta-correlated Gaussian process, we can write the Fokker-Plank equation
\begin{equation}
\label{FPE}
\partial_t P=s\partial_\varphi [\sin^2\varphi P]+2D\partial_\varphi^2P,
\end{equation}
in which $P(t,\varphi)$ is the probability density function of $\varphi$, and  $D$ denotes the angular diffusion coefficient.
Equation (\ref{FPE}) should be supplemented by the periodic boundary condition $P(t,-\pi/2)=P(t,\pi/2)$ and the normalization condition $\int_0^\pi P(t,\varphi)d\varphi=1$.
In the strong shear limit, $s\gg D$, the stationary solution of Eq. (\ref{FPE}) is localized at small angles and the steady-state average value of $\varphi$ is given by (see \citep{Tur07} for the details of derivation)
 \begin{equation}
 \label{phi_0}
 \langle \varphi\rangle\approx  \frac{3^{1/3}\sqrt{\pi}}{\Gamma(1/6)}\left(\frac{D}{s} \right)^{1/3}.
 \end{equation}
Here and in what follows, the angle brackets denote averaging over time or over realizations of the chaotic flow.


The near-wall shear rate can be estimated as $s\sim U_L/L$ where $U_L$ is the mean velocity of the flow at the border of the boundary layer.
Next, the angular diffusion coefficient is determined by the gradient of the wall-normal component of velocity fluctuations:  $D=\frac12\int_{-\infty}^{0}\langle\partial_xu_z(\boldsymbol{r}(t),t)\partial_xu_z(\boldsymbol{r}(0),0)\rangle dt$.
Using Eq. (\ref{zdep}) one obtains $\partial_xu_z\sim u_Lz^2/L^3$ where $u_L$ measures  the characteristic (wall-normal) fluctuating velocity at $z=L$, and, therefore, $D\sim(\partial_xu_z)^2\tau_c\sim u_L^2\tau_cz^4/L^6$.
Inserting the estimates of $s$ and $D$ into Eq. (\ref{phi_0}), we find the conditional average of $\varphi$ at a given distance $z$ from the wall
 \begin{equation}
 \label{phi_1}
 \langle \varphi\rangle\sim \frac{u_L^{2/3}\tau_c^{1/3}z^{4/3}}{U_L^{1/3}L^{5/3}}.
 \end{equation}
 
Next let us consider the dynamics of the polymer elongation $R$.
Since typical orientation angles are small, $\varphi,\theta\ll1$, we can replace $\sin\varphi$ in Eq. (\ref{R}) by $\varphi$, while $\cos\varphi$ and $\cos\theta$  can be replaced by unity.
Besides, we neglect the term $\varsigma_\parallel$ in comparison with $s\varphi$  because the diffusion coefficient $D_{\parallel}\sim (\partial_x u_x)^2\tau_c\propto z^2$ associated with $\varsigma_\parallel$ is much smaller than $s\langle \varphi\rangle\propto z^{4/3}$ at $z\ll L$.
Then in the main approximation the dynamics of $R$ is described by the following equation
\begin{equation}
\label{lnR}
\partial_t \ln R=-\gamma(R)+s\varphi.
\end{equation}
For the statistically stationary state we immediately find 
\begin{equation}
\label{f_02}
\langle \gamma\rangle= s\langle \varphi\rangle \sim \frac{u_L^{2/3}U_L^{2/3}\tau_c^{1/3}z^{4/3}}{L^{8/3}}.
\end{equation}

To proceed we need to relate the correlation time $\tau_c$ with the spatial scale $L$ and the characteristic velocities $U_L$ and $u_L$.
It is natural to expect that  $\tau_c$ is of the same order as the inverse Lyapunov exponent $\lambda_L^{-1}$ at the edge of the boundary layer.
Whereas the distance between two fluid parcels evolves accordingly to Eq. (\ref{lnR}) with $\gamma=0$, the Lyapunov exponent defined as the mean logarithmic rate of divergence of neighbouring Lagrangian trajectories is given by  $\lambda_L=s\langle\varphi\rangle$, where $\langle\varphi\rangle$ is taken at $z=L$.
Imposing the condition $\tau_c\sim \lambda_L^{-1}$, we find from Eq. (\ref{phi_1}) that $\tau_c\sim L/\sqrt{U_Lu_L}$. That yields the following estimates
\begin{equation}
\label{phi_2}
\langle \varphi\rangle\sim \left(\frac{u_L}{U_L}\right)^{1/2}\left(\frac{z}{L}\right)^{4/3},
\end{equation}
\begin{equation}
\label{f_2}
\langle \gamma\rangle\sim \frac{(u_LU_L)^{1/2}}{L}\left(\frac{z}{L}\right)^{4/3}.
\end{equation}
According to Eqs. (\ref{phi_2}) and (\ref{f_2}), the resulting effect of the hydrodynamic stretching can not be decomposed into the additive contributions of the stationary shear part and the fluctuating part of the flow. 
Instead, the polymer dynamics in the boundary layer is determined by the combined effect of regular and chaotic components: fluctuations of the velocity gradient provide polymer stretching in the wall-normal direction required for the polymer to feel the presence of the mean shear.


\section{Numerical simulations}

To check the scaling laws (\ref{phi_2}) and (\ref{f_2}) we performed the simulations with synthetic random flow which mimics the boundary layer of elastic turbulence. In the case of strong shear the stretching dynamics becomes essentially two-dimensional, so we restrict ourselves by 2d simulations. Namely, the components of the planar velocity field $(v_x,v_z)$ in the periodic domain $x\in [0,L]$, $z\in[-L,L]$  are chosen to be $v_x=U_x+u_x$, $v_z=u_z$, where
\begin{eqnarray}
\nonumber
U_x&=&\frac{U_L}{\pi}\sin \frac {\pi z}{L},\\
\label{synthetic_flow}
u_x&=&\frac{L^2}{\pi^2}\left[a_1(t) \cos \frac{2\pi x}{L} +a_2(t) \sin \frac{2\pi x}{L}\right] \sin \frac {\pi z}{L},\\
\nonumber
u_z&=&\frac{2L^2}{\pi^2}\left[a_1(t)\sin \frac{2\pi x}{L}-a_2(t) \cos \frac{2\pi x}{L}\right]
\left(1-\cos \frac{\pi z}{L}\right).
\end{eqnarray}
The independent random variables $a_1$ and $a_2$ are telegraph processes that is $a_1$ and $a_2$ remain constant during time slot $\tau_c$ and their values are chosen from identical normal distributions with zero mean and variances $\langle a_1^2 \rangle=\langle a_2^2\rangle=u_L^2/L^4$. Let us stress, that the velocity field $(v_x,v_z)$ is incompressible ($\partial_xv_x+\partial_zv_z=0$) and that it reproduces the proportionality laws $U_x,u_x\propto z$, $u_z\propto z^2$ when $|z|\ll L$.
In our numerics we chose $L=1$, $U_L=1$, $u_L=0.1$ and $\tau_c=L/\sqrt{U_Lu_L}$.
The polymer dynamics was modelled via two-dimensional version of equations (\ref{dynamics_r}) and (\ref{dynamics_R}) supplemented by the finitely extendible nonlinear elastic (FENE) model: $\gamma(R)=\gamma_0 (1-R^2/R_m^2)^{-1}$.  
The parameters $\gamma_0$, $R_m$, $\kappa$ and $\eta$ were  adjusted in such way that $R_m/R_0=10^2$, ${\it Wi}=10^4$ and ${\it Pe}=10^6$, 
where $R_0=\sqrt{3\eta/\gamma_0}$ is the radius of gyration of a polymer and  $\it Pe=U_LL/\kappa$ is the Peclet number.
 \begin{figure}
  \centerline{\includegraphics[scale=0.65]{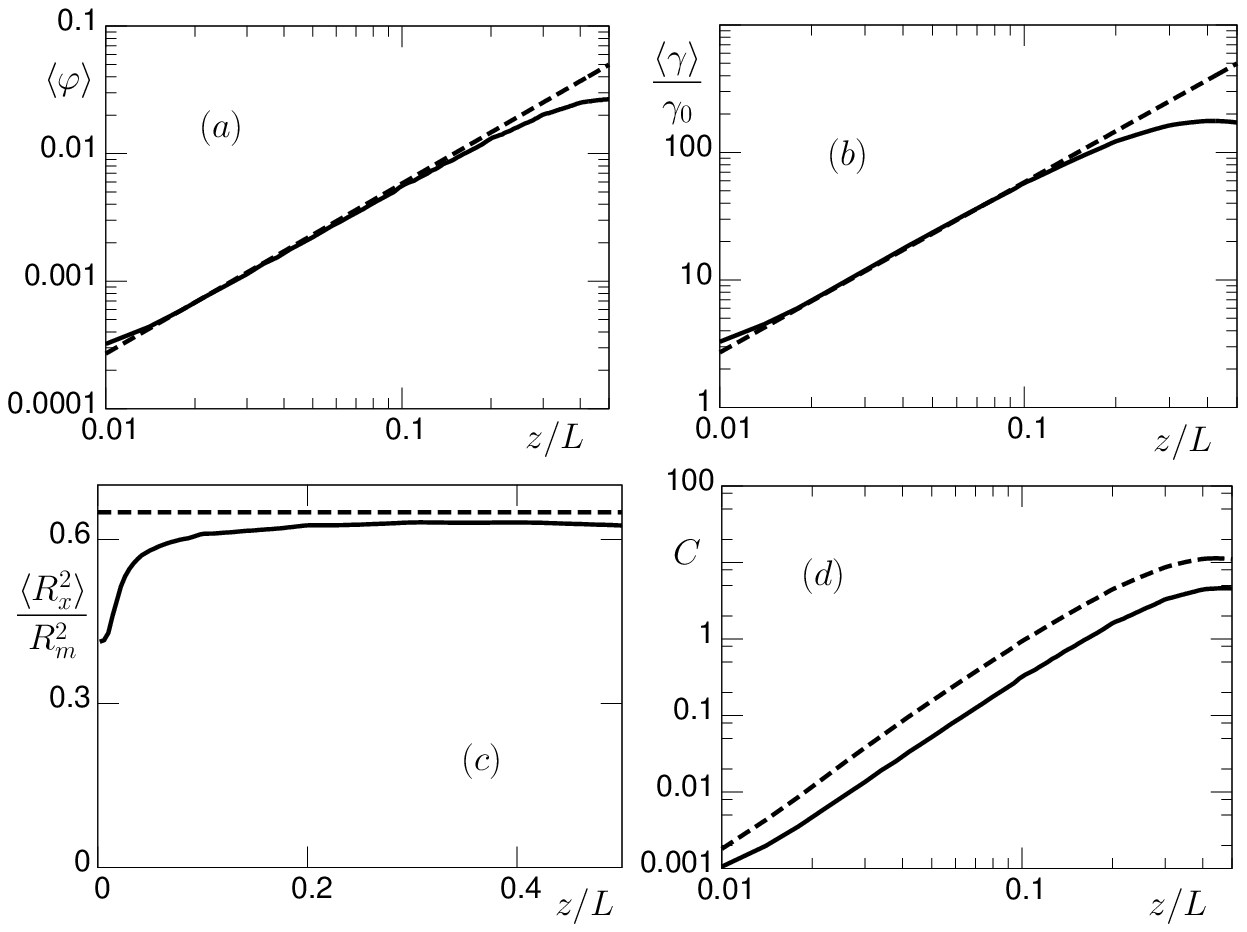}}
 \caption{(a) and (b): the average orientation angle $\langle \varphi\rangle$ and the spring constant  $\langle \gamma\rangle$  in dependence of the distance from the wall $z$. The solid lines represent the results of numerical simulations and the dashed lines are the theoretical power-law profiles, see Eqs. (\ref{phi_2}) and (\ref{f_2}).
(c): the mean square of the polymer elongation $\langle R_x^2\rangle$ along the shear direction as a function of distance from the wall $z$. The dashed line represents  the asymptotic value corresponding to the limit of infinitely large shear.   (d): the statistical moment $C=\langle \gamma R_xR_z\rangle$ (solid line) versus the distance from the wall $z$. This plot demonstrates that  $\langle\gamma R_xR_z \rangle$ depends on $z$ in the same manner as $\langle \gamma \rangle\langle \varphi \rangle R_m^2$ (dashed line).}
  \label{pic: numerics}
 \end{figure}
%
%

The upper panels of Fig. \ref{pic: numerics} show the steady-state expected values of the orientation angle  $\langle \varphi\rangle$ and spring constant $\langle \gamma\rangle$ in their dependence on $z$. In agreement with theoretical prediction, $\langle \varphi\rangle$ and $\langle \gamma\rangle$ exhibit power-law behavior with exponent $4/3$ starting from some small distance from the wall and till the scale of the order of $L$. The mean square of the polymer elongation along wall is approximately constant in this region, $\langle R_x^2\rangle\approx 0.62R_m^2$, see lower left panel of Fig. \ref{pic: numerics}. This is because $\langle R_x^2\rangle$ is close to its limiting value, corresponding to the limit of infinitely large shear, which is numerically found to be about $0.65R_m^2$ for the FENE model.
Note that the limiting value  differs from $R_m^2$ due to the ongoing tumbling of a polymer molecule \citep{Tur07}.

\section{Relation between mean flow and fluctuations}
Having described the single polymer statistics inside the boundary layer, we now pass to the macroscopic level of description.
The hydrodynamics of the low-$\Rey$ polymer solution is governed by interplay of viscous forces and elastic forces, while fluid inertia can be neglected. 
Namely,  if the flow is incompressible, i.e., the mass density of the neat fluid $\rho$ is a
constant and $\bnabla\bcdot\boldsymbol{v} = 0$,  the momentum conservation law (with the non-linear inertial term omitted)
reads 
\begin{eqnarray}
\label{Navier-Stokes}
\partial_t v_i &=& -\frac{1}{\rho}\nabla_i p +\nabla_j \Pi_{ij}^v+\nabla_j \Pi_{ij}^e.
\end{eqnarray}
Here $p$ is the pressure, 
$\Pi_{ij}^{v}$ is the viscous stress tensors, and $\Pi_{ij}^{e}$ is the tensor of elastic stresses  created by the polymer stretching in the flow. 
Equation (\ref{Navier-Stokes}) is applicable at scales much larger than the inter-polymer distance where the polymer solution can be regarded as a continuous medium.

%
%
%


The elastic stress is the source of instability leading to the turbulent-like state of a polymer solution. In the bulk region of elastic turbulence, where polymers provide strong back reaction on the velocity field, the elastic stresses are of the order or larger than the viscous stresses existing in the flow. However, in the peripheral region, the polymers are relatively weakly stretched since fluctuations of the velocity gradient diminish towards the non-slipping wall. This is why a boundary layer is formed, where viscosity overcomes elasticity.

%
%
%
%
%
%


For the statistically stationary flow, the only component of the viscous stress tensor remaining nonzero after averaging  is $\langle\Pi^{v}_{xz}\rangle=\nu \partial_z U_x$, where $\nu$ is Newtonian viscosity of the neat fluid. The corresponding component of the polymer-induced elastic stress is given by \citep{Bird}
\begin{equation}
\label{elastic_stress}
\langle\Pi_{xz}^e\rangle= \frac{nk_BT}{\rho\gamma_0}\frac{\langle \gamma(R)R_xR_z\rangle}{R_0^2},
\end{equation}
where $n$ is the concentration of the polymers in solution, 
$T$ is temperature, and $k_B$ is the Boltzmann constant.

It is difficult to derive a closed-form expression for the statistical moment $\langle\gamma  R_xR_z\rangle$ entering Eq.  (\ref{elastic_stress}). 
However, for the boundary-layer region one can safely write the estimate $\langle\gamma R_xR_z\rangle\sim \langle\gamma\rangle \langle \varphi\rangle R_m^2$, %
which is confirmed by numerical simulations, see lower right panel of Fig. \ref{pic: numerics}.
Then, using Eqs. (\ref{phi_2}), (\ref{f_2}) and (\ref{elastic_stress}) together with the relation $\gamma_0\sim k_BT/\rho \nu R_0^3$, we obtain $\langle \Pi^{e}_{xz}\rangle\sim \nu nR_m^2R_0u_Lz^{8/3}/L^{11/3}$ at $z\ll L$. By the definition of the boundary layer, the elastic stress must become of the order of the viscous stress $\langle\Pi^{v}_{xz}\rangle\sim\nu U_L/L$ at $z=L$. This condition produces the following relation
\begin{equation}
\label{turbulence_intensity}
\frac{U_L}{u_L}\sim nR_0R_m^2,
\end{equation}
which is the central result of this work.
Equation (\ref{turbulence_intensity}) shows that in contrast to high-$\Rey$ turbulence, where the fluctuating velocity is always of the order of the mean flow in the viscous sublayer, wall-bounded elastic turbulence possess a non-trivial relation between the mean and the fluctuating velocity components.

We are ready to justify the set of assumptions (i)-(iv) underlying the analytical procedure presented above.
First of all, we note that the characteristic time of the extensional dynamics is $\langle \gamma\rangle^{-1}$.
Then, as it follows from Eq. (\ref{f_2}) and from the estimate $\tau_c\sim L/\sqrt{U_Lu_L}$,  the assumption (i), which can be written as  $\tau_c\ll \langle \gamma\rangle^{-1}$, is self-consistent for any $z\ll L$.
There is a simple  explanation of why the time scales 
$\tau_c$ and $\langle \gamma\rangle^{-1}$  are well-separated inside the boundary layer.
In the bulk region of elastic turbulence, where fluid motion is strongly influenced by polymer back reaction, the velocity correlation time $\tau_c$ must be of the same order with the relaxation time of elastic stresses $\langle\gamma\rangle^{-1}$.
Since the flow fluctuations in the boundary layer are induced by bulk turbulence, then the correlation time $\tau_c$ near the wall is also determined by the value of $\langle \gamma\rangle^{-1}$ in the bulk.
At the same time, 
hydrodynamic stretching of polymers in close vicinity to the wall is much weaker than their stretching in the bulk.
We thus conclude that $\tau_c\ll \langle \gamma\rangle^{-1}$ in the boundary layer.

Next let us examine the range of validity for the ``adiabatic" assumption (ii). Since the center of mass of the polymer is passively advected by the flow, the probability density function $n(z,t)$ of its $z$-coordinate obeys the standard  advection-diffusion equation $\partial_t n=\partial_z[D_{zz}(z)\partial_z n]$, where $D_{zz}(z)=\mu z^4+\kappa$ and $\mu\sim u_L^2\tau_c/L^4$ \citep{LT_2004,Chernykh_2011}. The time required for the polymer at distance $z$ from the wall to ``feel" the inhomogeneity of fluctuations is estimated as $\tilde \tau(z)\sim z^2/D_{zz}(z)$. This time must be much larger than the characteristic time of the stretching dynamics, i.e. $\tilde{\tau}(z)\gg\langle \gamma\rangle^{-1}$.
Given the equation (\ref{f_2}), that leads to the following condition
\begin{equation}
\label{z_kappa}
z\gg r_{\kappa}=\left(\frac{U_L}{u_L}\right)^{3/20}\frac{L}{{\it Pe}^{3/10}}.
\end{equation}

%

As for the assumption (iii), the Langevin force $\boldsymbol{\zeta}$ produces the correction of the order of $\eta/R_m^2$ to the angular diffusion coefficient $D$. The correction is relatively small if
 \begin{equation}
 \label{z_eta}
 z\gg r_{\eta}=\left( \frac{R_0}{R_m}\right)^{1/2}\left(\frac{U_L}{u_L}\right)^{3/8}\frac{L}{{\it Wi}^{1/4}},
 \end{equation}
where ${\it Wi}=U_L\gamma_0^{-1}/L$ is the Weissenberg number .

Finally, the assumption (iv) is equivalent to the inequalities $\langle\varphi\rangle\ll1$ and $\langle \gamma\rangle\gg\gamma_0$.
The former is satisfied for any $z\ll L$ provided $u_L\lesssim U_L$, while the latter can be rewritten as
\begin{equation}
\label{z_gamma}
z\gg r_{\gamma_0}=\left(\frac{U_L}{u_L}\right)^{3/8}\frac{L}{{\it Wi}^{3/4}}.
\end{equation}

Taking into account Eq. (\ref{turbulence_intensity}), we conclude from Eqs. (\ref{z_kappa})-(\ref{z_gamma}) that the spatial scales $r_{\kappa}$, $r_{\eta}$ and $r_{\gamma_0}$ are small in comparison with the width of the boundary layer $L$ in the limit when both the Weissenberg number $\it Wi$ and the Peclet number $\it Pe$ are large.
That means the self-consistency of our asymptotic theory.

\section{Discussion}

 We have combined a microscopic description of the polymer statistics with a macroscopic description of the stresses in a polymer solution to investigate the boundary-layer properties of elastic turbulence.
That allowed us to derive the parametric relation between the magnitude of the velocity fluctuations and the mean
flow in the high Weissenberg number limit, see Eq. (\ref{turbulence_intensity}). 
This relation follows from the condition that the elastic stresses become comparable with the viscous stresses at the border of the boundary layer.

Unfortunately, the experimental studies performed to date are beyond applicability of the asymptotic analysis presented here.
Say, for the typical experimental parameters $nR_0^3\approx 0.1$ and $R_m/R_0\approx 100$ one observes $U_L/u_L\approx10$ and, therefore, we need to take ${\it Wi}\gtrsim 10^4$  to ensure the spatial scale separation $r_{\kappa}, r_{\eta},r_{\gamma_0} \ll z\ll L$.
At the same time, the maximum Weissenberg number in existing experiments is of the order of $10^3$.
Moreover, the large values of Weissenberg number, which we need to justify our analytical scheme, are difficult to achieve in practice because of too fast mechanical degradation of polymer molecules at high shear rates.
Thus, the current lack of relevant experimental data does not allow us to test the scaling laws (\ref{phi_2}) and (\ref{f_2}) and the parametric dependence predicted by Eq. (\ref{turbulence_intensity}).

Could the asymptotic results give us qualitatively correct insight into the boundary layer properties outside the relevant asymptotic regime?
Although the systematic study of this issue requires much further experimental work, the already known  data suggest that Eq. (\ref{turbulence_intensity}) properly captures the observed peculiarities of elastic turbulence.
First of all, it is noteworthy that the boundary layer width $L$ drops from Eq. (\ref{turbulence_intensity}).
This is in accord with the experimental observation by \cite{Jun_2011} that $L$ is determined by the system size rather than being merely a function of the properties of a polymer solution.
Secondly, velocity fluctuations are always weaker than the average flow in the elastic turbulence experiments.
In qualitative agreement with this fact, our theory predicts that at $nR_0^3\approx 0.1$ and $R_m/R_0\approx 100$  the mean velocity is
large compared to the fluctuating velocity  provided the unknown numerical prefactor in the right hand side of  Eq. (\ref{turbulence_intensity}) is not too small.
Finally, and most importantly, Eq. (\ref{turbulence_intensity}) tells us that fluctuations become less pronounced with increase of the polymer concentration.
This result may seem to be somewhat paradoxical: although the phenomenon of elastic turbulence becomes possible due to polymer additives, the turbulence intensity decreases as the concentration of the additives grows.
However, the data of the recent experimental study of elastic turbulence in a von
Karman swirling flow speak in support this prediction, see \cite{Jun_2017}.
We hope that these remarks will stimulate the detailed experimental and numerical verification of the theoretical results presented here.


We are grateful to V. Steinberg for valuable discussions.
This work was supported by the Russian Scientific
Foundation, Grant No. 14-22-00259.

\bibliographystyle{jfm}

{}

\end{document}